\documentclass[useAMS,usenatbib]{mn2e}
\usepackage{amssymb}
\usepackage{graphics}
\usepackage{tabularx}
\usepackage{color}

\def\eqref#1{equation~(\ref{#1})}

\title[HAT-P-13b and TTVs]{Transit timing variations in the HAT-P-13 planetary system}
\author[P\'al et al.]{%
Andr\'as P\'al$^{1,2}$%
\thanks{E-mail address: apal@szofi.net}\thanks{Bolyai Fellow},
Kriszti\'an S\'arneczky$^{1}$, 
Gyula M. Szab\'o$^{1,3}$,
Attila Szing$^{1}$, 
\newauthor
L\'aszl\'o L. Kiss$^{1,4}$, 
Gy\"orgy Mez\H{o}$^{1}$ and
Zsolt Reg\'aly$^{1}$ \\
$^{1}$ Konkoly Observatory of the Hungarian Academy of Sciences, 
        Konkoly Thege Mikl\'os \'ut 15-17,
        Budapest, H-1121, Hungary, \\
$^{2}$ Department of Astronomy, Lor\'and E\"otv\"os University, 
	P\'azm\'any P\'eter s\'et\'any 1/A, 
	Budapest H-1117, Hungary, \\
$^{3}$ Department of Experimental Physics and Astronomical Observatory, 
	University of Szeged, 
	6720 Szeged, Hungary \\
$^{4}$ Sydney Institute for Astronomy, 
	School of Physics A28, University of Sydney, 
	NSW 2006, Australia}

\begin{document}

\date{Accepted ..., Received ... ; in original form ...}

\pagerange{\pageref{firstpage}--\pageref{lastpage}} \pubyear{2010}

\maketitle

\label{firstpage}

\begin{abstract}
In this Letter we present observations 
of recent HAT-P-13b transits. The combined analysis of published 
and newly obtained transit epochs shows evidence
for significant transit timing variations
since the last publicly available ephemerides. 
Variation of transit timings result in a sudden switch of transit
times. The detected full range of TTV spans $\approx0.015$\,days, which
is significantly more than the known TTV events exhibited by hot
Jupiters. If we have detected a periodic process, its period should be
at least $\approx 3$\,years because there are no signs of variations in
the previous observations. This argument makes unlikely that the
measured TTV is due to perturbations by HAT-P-13c.
\end{abstract}



\section{Introduction}
\label{sec:introduction}

Transit timing variations (TTVs) of transits are expected in transiting extrasolar 
planetary systems where the host star has more than one companion. 
This effect might reveal the presence of other companions with smaller masses
\citep[see e.g.][]{agol2005,steffen2007} or co-orbital bodies as 
well \citep{ford2007}. On the other hand, if the system has two or more
known planetary components, the magnitude of the TTV effects can be
predicted and/or used to characterize the system more precisely
\citep{holman2010,steffen2010}. Similarly, the lack of TTVs can rule out 
other companions above a certain limit \citep{csizmadia2010}. 
One should note that even two-body systems
(i.e. the parent star and the planet) can show significant TTVs
due to non-gravitational physical processes \citep[see e.g.][]{fabrycky2008}
and in the case of eccentric orbits, there are intrinsic timing variations
due to the effects of general relativity \citep[see e.g.][]{pal2008b}.
In the case of two planetary companions, the magnitude of the TTV effects 
can be predicted analytically using the same methodology that is
known for hierarchical stellar systems \citep{borkovits2003}, 
and of course, by numerical integrations. 

At the time of its discovery, the planetary system orbiting
the star HAT-P-13 was the only known multiple extrasolar planetary
system that exhibits at least one transiting planet as well 
\citep[HAT-P-13b,][]{bakos2009}.
Follow-up studies showed that the host star has a spin alignment
with respect to the orbital plane of the transiting planet \citep{winn2010}
as well as this work also noticed a significant long-term drift
in the radial velocity of the host star that might be interpreted
as a presence of a third companion. 
Additional studies \citep{szabo2010} reported the lack of 
primary transits of the long-period second companion HAT-P-13c,
and presented additional observations from the transits of
the close-in planet as well. As of this writing, no other 
\citep[than][]{bakos2009,szabo2010} publicly 
available photometric data are available for this planetary system. 
Here we describe recent photometric measurements of 
the host star. The analysis of the light curves showed significant
timing variations  since the last public ephemerides.
The structure of this Letter is as follows. In Section~\ref{sec:observations}
we describe details of the observations, data reduction and light 
curve modelling. The analysis of the results yielded by light curve fits 
is presented in Section~\ref{sec:timing}. Section~\ref{sec:discussion}
discusses possible scenarios for the observed
TTVs and summarizes our results.


\begin{figure}
\begin{center}
\resizebox{80mm}{!}{\includegraphics{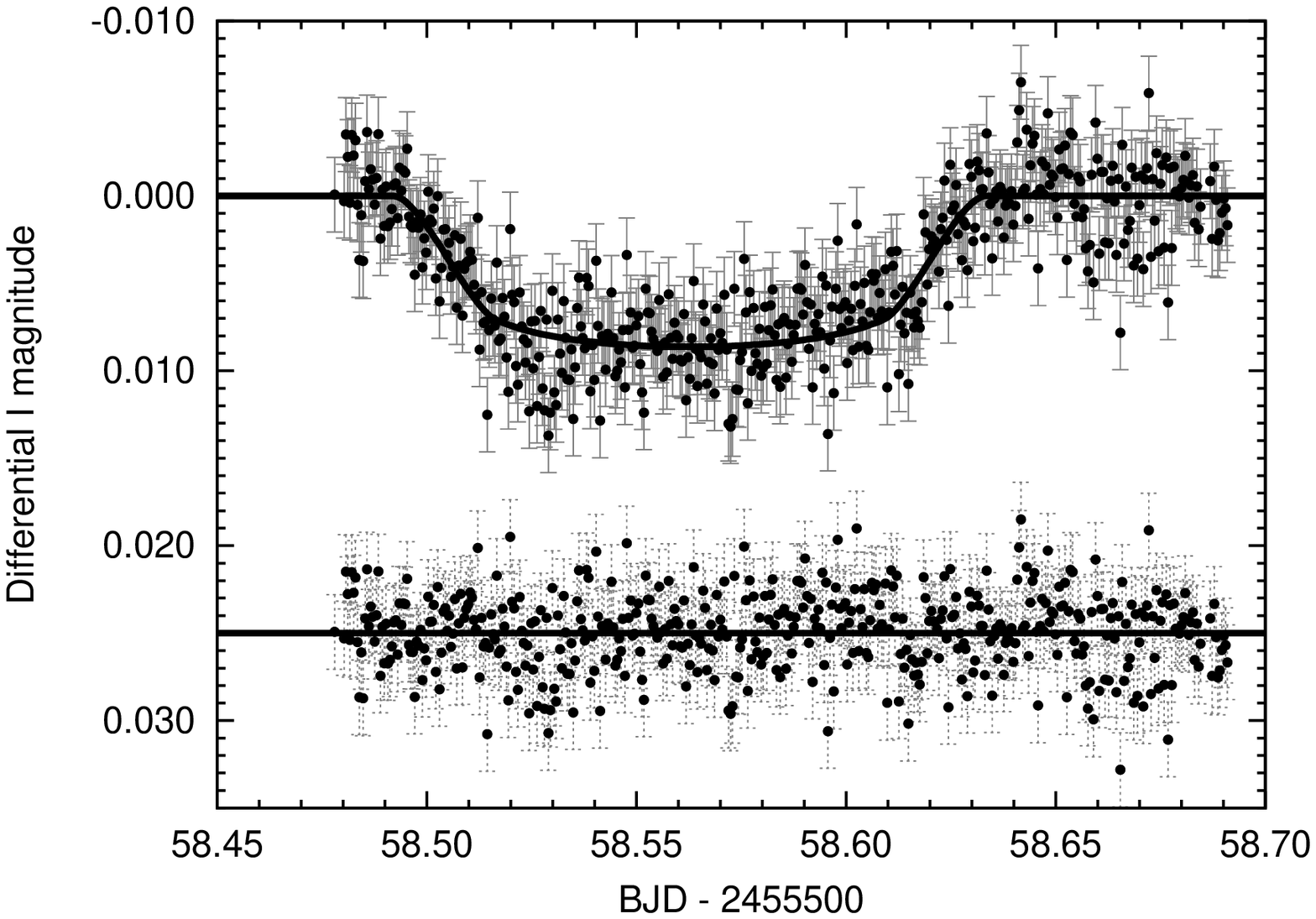}}
\resizebox{80mm}{!}{\includegraphics{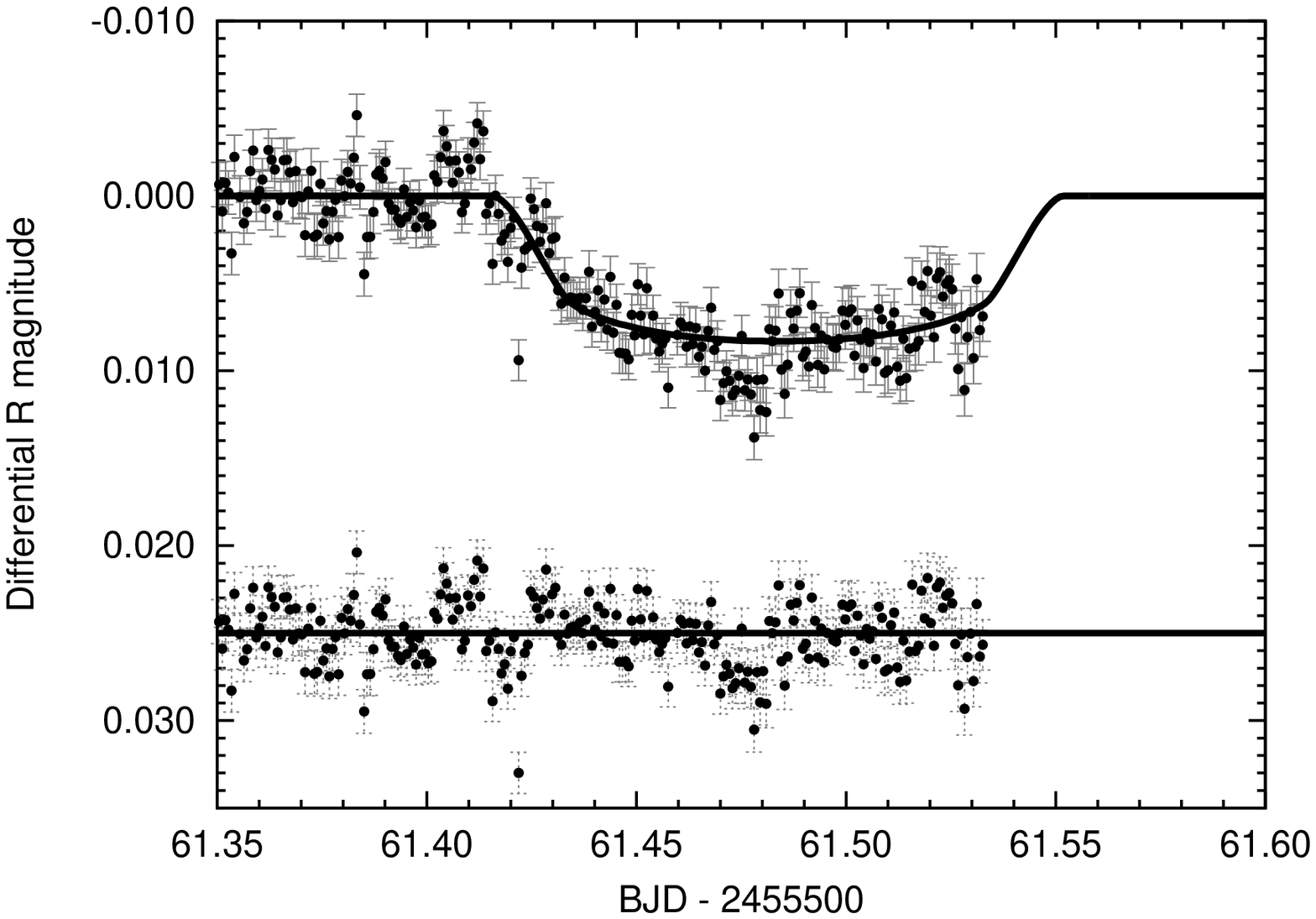}}
\resizebox{80mm}{!}{\includegraphics{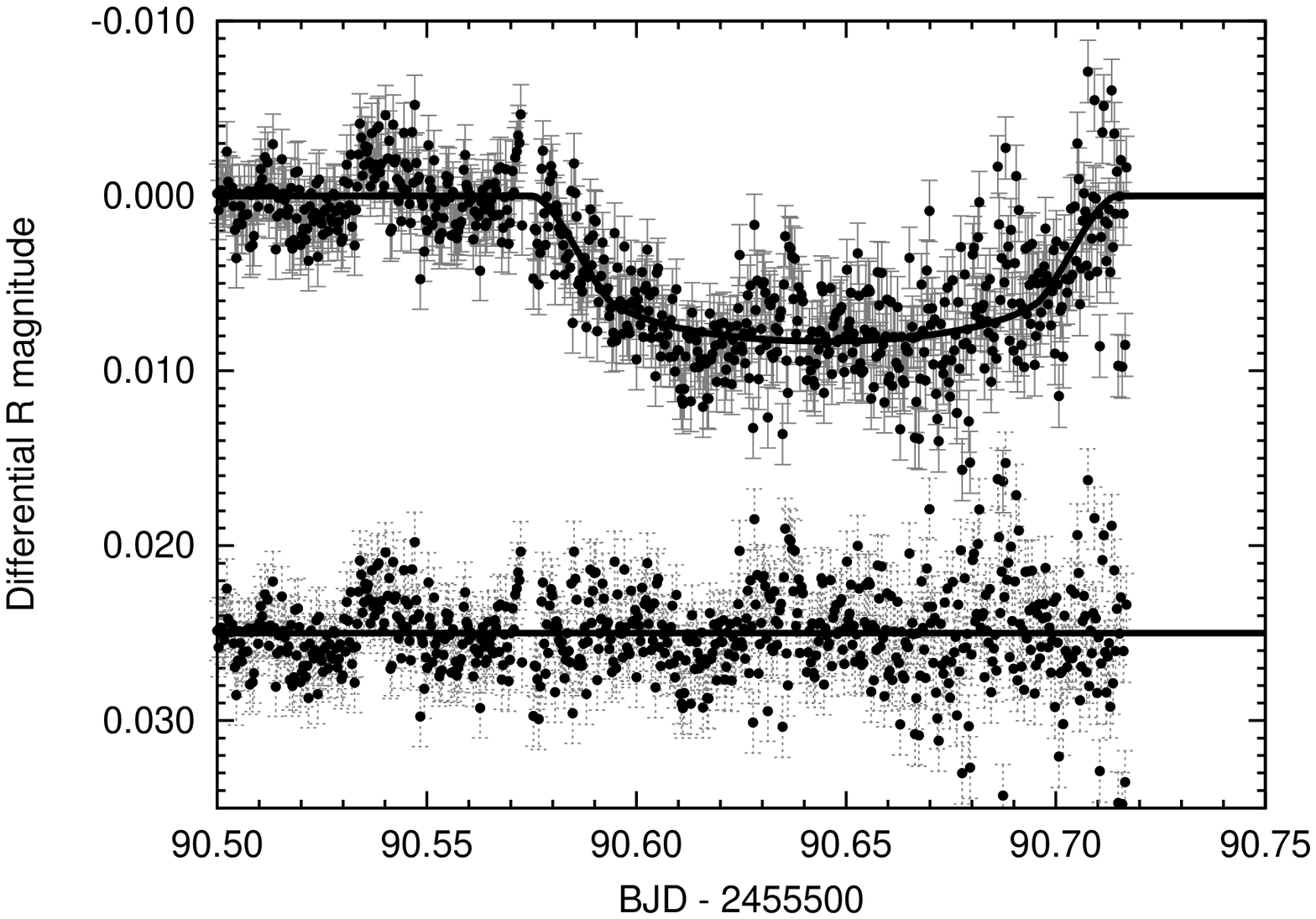}}
\end{center}
\caption{Photometry of the star HAT-P-13 on the nights of
2010 December 27/28 (upper panel), December 30/31 (middle panel)
and 2011 Januar 28/29 (lower panel),
using the facilities of the 
Piszk\'estet\H{o} Mountain Station, Konkoly Observatory.
Superimposed are the best-fit transit model light curves. Below
the light curve plots, the fit residuals are also shown. See
text for further details.}
\label{fig:hatp13-lc}
\end{figure}

\section{Observations and data reduction}
\label{sec:observations}

We carried out photometric observations of the star HAT-P-13 with
the telescopes of the Konkoly Observatory, located at the 
Piszk\'estet\H{o} Mountain Station on three separate nights. 
On the night of 2010 December 27/28, we used the 60/90/180\,cm
Schmidt telescope, equipped with an Apogee ALTA-U $4{\rm k}\times 4{\rm k}$ 
CCD camera.
The following two events, on the night of December 30/31 and 2011 
Januar 28/29 has been partially observed (due to weather circumstances
on December 30 and due to dawn twilight on Januar 28)
with the 100\,cm Ritchey-Chr\'etien-Coud\'e (RCC)
telescope equipped with a Princeton Instruments VersArray $1.3{\rm k}\times 1.3{\rm k}$ 
camera. These setups of optics and cameras yielded a square-shaped
field-of-view with a size of $1.17$ and $0.11$ degrees, for the
Schmidt and RCC telescopes, respectively. 
At the first night we used Bessel I filter and an exposure time of 
20\,seconds (42\,seconds cadence) while in the second and third night we
used Cousins R filter and an exposure time of 60 and 25\,seconds, respectively
(here the readout time is relatively short, so this can be used as a 
cadence as well and on the last night the seeing was much better hence the
shorter exposure time). 

In the following we describe the data reduction processes and some
aspects of the light curve modelling.

\subsection{Data reduction}

Reduction of raw technical and scientific frames has been carried
out with the pipelines built on the software package described
in \cite{pal2009}. The calibration procedure followed the standard
way of dark subtraction and flat-field corrections using 
dark frames taken with similar exposures as the scientific images.
(employing the tasks \texttt{ficalib} and \texttt{ficombine}). 
The source identification, astrometric solution and centroid coordinates
were derived as follows. First, individual point-like sources were
extracted from each frame using the task \texttt{fistar}. These lists 
of profile coordinates were then cross-matched \citep{pal2006}
between the frames
and a reference frame,
yielding both a list of matched pairs and the spatial transformation
between two images. The polynomial order of the spatial transformation
were 3 and 1 for the Schmidt and RCC fields, respectively (increasing
these orders did not decrease the unbiased fit residuals). Then,
using these initial differential astrometric solutions, the stellar pixel
coordinates were refined using an independent profile centroid fit
(done by the task \texttt{fiphot})
for the expected positions. These lists of refined profile coordinates
were used to refine the spatial transformation as well. This is a rather
relevant step because the point matching task (\texttt{grmatch}) sometimes 
misses a point and the exclusion of even a single point can yield a
systematic deviation in the transformation coefficients whose error
can propagate into the photometry errors (resulting in unexpected red noise). 
Following this two-stage astrometry,
these refined coefficients were used to transform the reference object 
list to the system of individual frames to perform aperture photometry. 
Involving the task \texttt{fiphot}, aperture photometry was performed 
with annuli of appropriate sizes and
a series of apertures from which we picked the best one later on the 
modelling. Differential magnitudes were then computed from these
raw instrumental magnitudes using several nearby comparison stars. The
white-noise uncertainties of the individual photometric points were
derived from the photon noise, background noise and scintillation 
\citep{young1967}
noise components of both the target star and comparison stars. 

\subsection{Light curve modelling}

Light curves obtained by the procedure (as described above) have been
used to model the transit event and derive the respective parameters.
In order to fit a model light curve to the observed data, we employed
the analytic formulae of \cite{mandel2002}. The effects of 
stellar limb darkening have been taken into account by fixing the 
quadratic limb darkening coefficients using the stellar atmospheric
parameters \citep{bakos2009}. These coefficients have been derived from
the tables of \cite{claret2000}. Since many of the regression methods
used by us require the knowledge of parametric derivatives of the model
function, we used the formulae provided by \cite{pal2008} to compute these
partial derivatives.

The light curve models have been parameterised using 
$p\equiv R_{\rm p}/R_\star$, the planet-to-star size ratio, 
$b^2$, the square of the impact parameter,
the quantity $\zeta/R_\star$, related to the duration of the transit
as $\zeta/R_\star=2/(T_{3.5}-T_{1.5})$ and 
$T_{\rm c}=(T_{1.5}+T_{3.5})/2$, the time instance at the center of the 
transit (here $T_{1.5}$ and $T_{3.5}$ denotes the time instances
when the center of the planet disc intersects the limb of the star inwards
and outwards, respectively). See also \cite{pal2008} for further details
on the advantages of this parameterisation. 
In addition
to these parameters, we fitted simultaneously the out-of-transit magnitude,
a linear trend in the out-of-transit magnitude (correcting to the effects
of changing airmasses) and a linear decorrelation parameter against 
the variations resulted by the changes in the profile FWHM. In
the case of the observation on 2010 December 30/31, we kept 
$R_{\rm p}/R_\star$, $b^2$ and $\zeta/R_\star$ fixed, since this
light curve does not cover the entire transit event. We employed 
the Markov Chain Monte Carlo method \citep[MCMC; see][]{ford2004} to
obtain the \emph{a posteriori} distribution of the fit parameters
from which the best fit values, uncertainties and correlations can 
easily be derived. The 
best fit values for the light curve shape parameters for 
the event 2010 December 27/28 are 
$R_{\rm p}/R_\star=0.0880\pm0.0031$, 
$b^2=0.544\pm0.086$ and 
$\zeta/R_\star=17.07\pm0.28$. These values agrees well with the
previously published ones \citep{bakos2009}.
Fig.~\ref{fig:hatp13-lc} shows the light curves with these
best fit model functions superimposed while 
the best fit values for the transit 
center instances are listed in Table~\ref{table:timing}.
This table also lists the previously published individual transit event
instances as well, from \cite{bakos2009} and \cite{szabo2010}. 
The fit residuals has been compared to the expected residual value
that can be computed from the photon noise, background noise and
scintillation of the star. We find that the fit residuals
for the observation of 2010 December 27/28 were only larger by
a factor of $1.05$, that indicates a very good quality light curve
with small systematic variations (red noise). Indeed, even the
raw instrumental magnitudes (i.e. the magnitude of the target 
star without subtracting any comparison star) varied in a range
of only $0.02$\,mag. However, the analysis of the
observation from 2010 December 30/31 yielded a fit residual that is
larger by a factor of $\approx 2$ than the expected and as it 
clear from the lower plot on Fig.~\ref{fig:hatp13-lc}, the residuals 
contain significant amount of red noise. This was due to both weather
circumstances and significant flat field image structures in the
vicinity of HAT-P-13. Therefore we conservatively scaled up the
derived errors by that factor of $2$ (and this is the value 
that has been reported in Table~\ref{table:timing}).
The observation carried out on 2011 January 28/29 was partial:
the out-of-transit part of the light curve after the transit
was not measured, however, the observed egress can be incorporated into
the light curve analysis. This fit has been done similarly as for the night
of December 27/28 and yielded 
$R_{\rm p}/R_\star=0.0820\pm0.0078$, 
$b^2=0.471\pm0.214$ and 
$\zeta/R_\star=16.88\pm0.48$. These values also agrees well (within
uncertainties) both with the fit of December 27/28 data and the 
published values. For the center of the transit, we obtain the value
found in Table~\ref{table:timing}. For this analysis, the uncertainties 
yielded by the MCMC fit have been multiplied by $1.5$ in order to
take into account the red noise components: this value is the
ratio of the fit residuals and the expected formal photometric 
errors (those have been derived from the photon and background noise 
and the scintillation). The uncertainties reported above for
the geometric parameters and listed in Table~\ref{table:timing} are these
increased values.


\begin{table}
\caption{Transit center instances from previous
works and recent measurements. Data for the events $-68$ \dots $+62$
are available from Bakos et al. (2009), for the events $124$ and $161$
data are taken from Szab\'o et al. (2010). The last three fields are the
new measurements presented in this Letter. The event enumeration is
from the same reference epoch as in Bakos et al.~(2009). 
See text for further details.}
\label{table:timing}
\begin{center}\begin{tabular}{rlr}
\hline
Event 	& BJD	& Error \\
\hline
$-68$  	& 2454581.62406	&    0.00122 \\
$-1$   	& 2454777.01287	&    0.00100 \\
$0$     & 2454779.92953	&    0.00063 \\
$1$     & 2454782.84357	&    0.00155 \\
$24$    & 2454849.92062	&    0.00075 \\
$35$    & 2454882.00041	&    0.00150 \\
$62$    & 2454960.73968	&    0.00178 \\
$124$   & 2455141.55220	&    0.00100 \\
$161$   & 2455249.45080	&    0.00200 \\
$267$   & 2455558.56265	&    0.00098 \\
$268$   & 2455561.48379	&    0.00400 \\
$278$   & 2455590.64486 &    0.00179 \\
\hline
\end{tabular}\end{center}
\end{table}

\section{Timing variations}
\label{sec:timing}

\begin{figure}
\begin{center}
\resizebox{80mm}{!}{\includegraphics{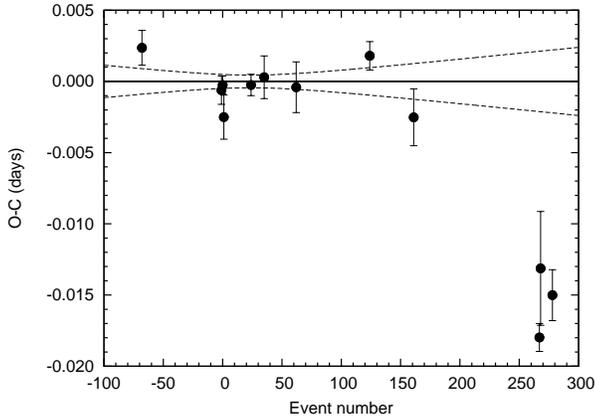}}
\end{center}
\caption{Transit timing variations as the function of the event number
(cycles since event at the night of 2008 November 9/10). The points
with the errorbars represent the individual measurements and 
the respective uncertainties as listed in
Table.~\ref{table:timing}. The thick solid line is the best fit
linear ephemerides to the first 9 data points while the thinner
dashed lines above and below it represents the 1-$\sigma$ errorbars
of the best linear fit. }
\label{fig:hatp13-ttv}
\end{figure}

Using the data of \cite{bakos2009} and \cite{szabo2010}, the subsequent
observed transit events can be modelled easily with a strictly
periodic assumption: the formal errors reported in these papers
yield an unbiased residual of $\chi_0=9.2$ for these 7 degrees 
of freedom. By defining the event $N_{\rm tr}=0$ as the 
reference epoch, this linear model yields $E=2454779.92978\pm0.00049$ (BJD)
and $P=2.9162953\pm0.0000085$~days while ${\rm Corr}(E,P)=-0.406$.
These ephemerides gives an uncertainty of $\Delta t=0.0021$~days
for the predictions of the discussed events ($N_{\rm tr}=267$, $268$ and
$278$ for the nights 2010 December 27/28, 30/31 and January 28/29).
As it is clear from Fig.~\ref{fig:hatp13-ttv}, these three recent measurements
outlie from the linear fit by $8.4$-$\sigma$, $3.3$-$\sigma$ and 
$5.5$-$\sigma$, respectively. 
We note here that the significance of the first (December 27/28)
measurement is determined mostly by the uncertainty of the predictions
while the significance of the second one is determined by the 
uncertainty of that particular measurement. All in all, 
both values can be treated as a significant deviation.

Due to the significance of the results, i.e. the deviance of the
transit center instances from the predictions, we applied several
independent analysis methods. These methods include the
minimisation algorithm based on the downhill simplex method
\citep{press1992}, as well as a one-parameter scan. In the latter test, the
light curve shape parameters ($R_{\rm p}/R_\star$, $b^2$ and $\zeta/R_\star$)
have been fixed and only the $T_{\rm c}$ has been varied while the
out-of-transit magnitude and other decorrelation parameters (see above)
have been minimised using the classic linear least squares method.
Additionally, we compared the statistical covariances with the
covariances yielded by linear error propagation analysis.
All of these methods have confirmed the results and uncertainties
of the Markov Chain Monte Carlo analysis presented in the previous section. 
In addition to these tests, we checked other data acquistion logs
to exclude other sources of possible systematic effects. The
filesystem metadata information has been compared with the FITS keywords
as well as the logs of the NTP daemon were also checked (this showed
an approximately 0.002~seconds of standard deviation from the time 
synchronization servers).


\section{Discussion}
\label{sec:discussion}

There are a few examples for extrasolar planets proven to exhibit TTV.
WASP-3b has been found to display a full TTV amplitude of $0.004$~day
period. As an explanation, a third planet has been inferred on a
resonance orbit that may be either an inner or outer perturber
\citep{maciejewski2010a}. The TTV of WASP-5 is
about 0.0023 day, and a set of different perturber solutions were
invoked that can equivalently well explain the observed TTV \citep{fukui2010}.
WASP-10 exhibits a similar TTV to that
of WASP-3 and WASP-5, its TTV has a full amplitude of $0.002$~day with
$5.47$\,day period. The perturber has been suggested to have $5\,M_{\rm J}$ 
mass and orbit in $5:3$ mean motion resonance with WASP-10b 
\citep{maciejewski2010b}. Other sources than perturbing planets are also 
known as sources of TTV, such as exomoons orbiting the 
planets \citep{szabo2006,simon2007}. However, the transit timing 
variations are in the order of a few seconds/minutes in the case of 
plausible moons, and the measured TTV of HAT-P-13b is not compatible 
with this predicted TTV amplitude.

The detected $O-C$ of HAT-P-13b is in the order of $0.01$~days, 
significantly more than
that of WASP-3, WASP-5 or WASP-10. Kepler-9b,c represent an example for
such large TTV amplitude exhibited by Neptunes that orbit in strong
resonance \citep{holman2010}. However,
HAT-P-13b is not very similar to the case of Kepler-9b,c. The planets
in the Kepler-9 system exhibit a continuous variation of TTV with a
period of few hundred days, while HAT-P-13b seemed to lack TTV during
the first three years of follow-up observations, and then a sudden
switch of TTV is observed.

This observed behaviour and the TTV amplitude of HAT-P-13b may be
explained by perturbations by long period planets on non-resonant
orbits. For instance, in \cite{borkovits2010}, a model planet
similar to CoRoT-9 exhibits a large switch in $O-C$ of the inner planet
in less than $1$\,year. Assuming an outer perturber of $5\,M_{\rm J}$ mass,
$10,000$~day orbital period and $e=0.7$ eccentricity, the predicted
amplitude results to be $0.013$~days (see Fig.~6 in the cited paper).
This value is quite similar to the $O-C$ variation that we report for
HAT-P-13, suggesting that perturbations on the time-scale of the
orbital period of a long-period perturber can explain our inferred TTV
of HAT-P-13b. The detailed modeling is beyond the information content
of the currently available dataset: there would be at least $6$ orbital
parameters to explain with the amplitude and the timing of the switch
of the $O-C$ diagram, leading to a highly unconstrained problem. It may
be unlikely that we observed perturbations of HAT-P-13c, because this
non-resonant perturber would lead to TTV variations with $428.5$~day
amplitude and additional secular components. This period is covered by
the current dataset and there are no signs for this period in the
deduced TTV. Eventually, the most urgent task is deriving the period of the
TTV shown by HAT-P-13b before predictions can be given to the
nature of the perturber.


\section*{Acknowledgments}

The work of A.~P. has been supported by the ESA grant PECS~98073
and by the J\'anos Bolyai Research Scholarship of the 
Hungarian Academy of Sciences. This project has also been supported by 
the ``Lend\"ulet'' Young Researchers Program of the Hungarian Academy 
of Sciences and the Hungarian OTKA Grants K76816, K83790, and MB08C 81013.


{}

\bsp

\label{lastpage}

\end{document}